\documentclass{IEEEcsmag}

\usepackage{graphicx}
\usepackage{xspace}
\usepackage{xcolor}
\usepackage[warn]{textcomp}
\usepackage{cite}
\usepackage[colorlinks,urlcolor=blue,linkcolor=blue,citecolor=blue]{hyperref}

\jvol{XX}
\jnum{XX}
\paper{X}
\jmonth{January/March}
\jname{Pervasive Computing}
\pubyear{2021}

\makeatletter
\def\ps@IEEEtitlepagestyle{%
  \def\@oddfoot{\mycopyrightnotice}%
  \def\@evenfoot{}%
}
\def\mycopyrightnotice{%
  { \footnotesize \url{https://doi.org/10.1109/MPRV.2020.3029650}\\ \copyright 2021 IEEE.  Personal use of this material is permitted.  Permission from IEEE must be obtained for all other uses, in any current or future media, including reprinting/republishing this material for advertising or promotional purposes, creating new collective works, for resale or redistribution to servers or lists, or reuse of any copyrighted component of this work in other works\hfill\vfill}
}
\makeatother

\begin{document}
\sptitle{Special Issue on Pervasive Manufacturing}
\editor{Florian Michahelles, Nadya Peek, and Simon Mayer}

\title{The Road to Ubiquitous Personal Fabrication: Modeling-free Instead of Increasingly Simple}

\author{Evgeny Stemasov}
\affil{Institute of Media Informatics, Ulm University, Ulm, Germany}

\author{Enrico Rukzio}
\affil{Institute of Media Informatics, Ulm University, Ulm, Germany}

\author{Jan Gugenheimer}
\affil{Télécom Paris - LTCI, Institut Polytechnique de Paris, Paris, France}

\markboth{}{The Road to Ubiquitous Personal Fabrication}

\begin{abstract}
    The tools for personal digital fabrication (DF) are on the verge of reaching mass-adoption beyond technology enthusiasts, empowering consumers to fabricate personalized artifacts.
	We argue that to achieve similar outreach and impact as personal computing, personal fabrication research may have to venture beyond ever-simpler interfaces for creation, towards lowest-effort workflows for remixing.
	We surveyed novice-friendly DF workflows from the perspective of HCI.
	Through this survey, we found two distinct approaches for this challenge: 1)  simplifying expert modeling tools (AutoCAD \textrightarrow Tinkercad), 2) enriching tools not involving primitive-based modeling with powerful customization (e.g., Thingiverse).
	Drawing parallels to content creation domains like photography, we argue that the bulk of content is created via remixing (2).
	In this work, we argue that to be able to include the majority of the population in DF, research should embrace omission of workflow steps, shifting towards automation, remixing, and templates, instead of modeling from the ground up.
\end{abstract}

\maketitle

\mycopyrightnotice

    \label{sec:introduction}
    \chapterinitial{Personal fabrication} (PF) describes the notion that machinery, workflows, and tools for industrial manufacturing become available to consumers.
    This -- ideally -- includes not only technology enthusiasts, but also less ''tech-savvy'' users.
    They may still desire to benefit from the opportunities of PF, such as tailored artifacts they are unable to order online easily (e.g., non-standardized attachments~\cite{baudischPersonalFabrication2017,chenRepriseDesignTool2016,shewbridgeEverydayMakingIdentifying2014}).
    However, these potential users may not be convinced to invest time in skill acquisition and PF processes. %
    They may not be ''makers'' and may likewise not be convinced to learn digital fabrication (DF) for their benefit -- especially when their alternatives are ever-improving online shopping ''workflows''.\\
    
    Just like computing itself progressed from centralized use for few, often expert, user groups to personal and ubiquitous computing, (personal) fabrication is likely on a similar path.
    Ultimately, as described by Gershenfeld, we may have machines able to fabricate \emph{anything}~\cite{gershenfeldDesigningRealityHow2017}.
    Such a device is still constrained by the input it may receive -- i.e., ''what to fabricate?'', currently answered by the use of (computer-aided) design software. 
    We want to approach these developments from the perspective of HCI. 
    Namely, the assumption that machines able to fabricate in any material and size will be available to users in the same way powerful word, video, and image processors became available to and actively used by them.
    This empowers users of PF devices to benefit from digital precision to create and shape matter -- both for productive and mundane purposes.
     
    \begin{figure*}[ht!]
        \centering
        \includegraphics[width=0.9\linewidth]{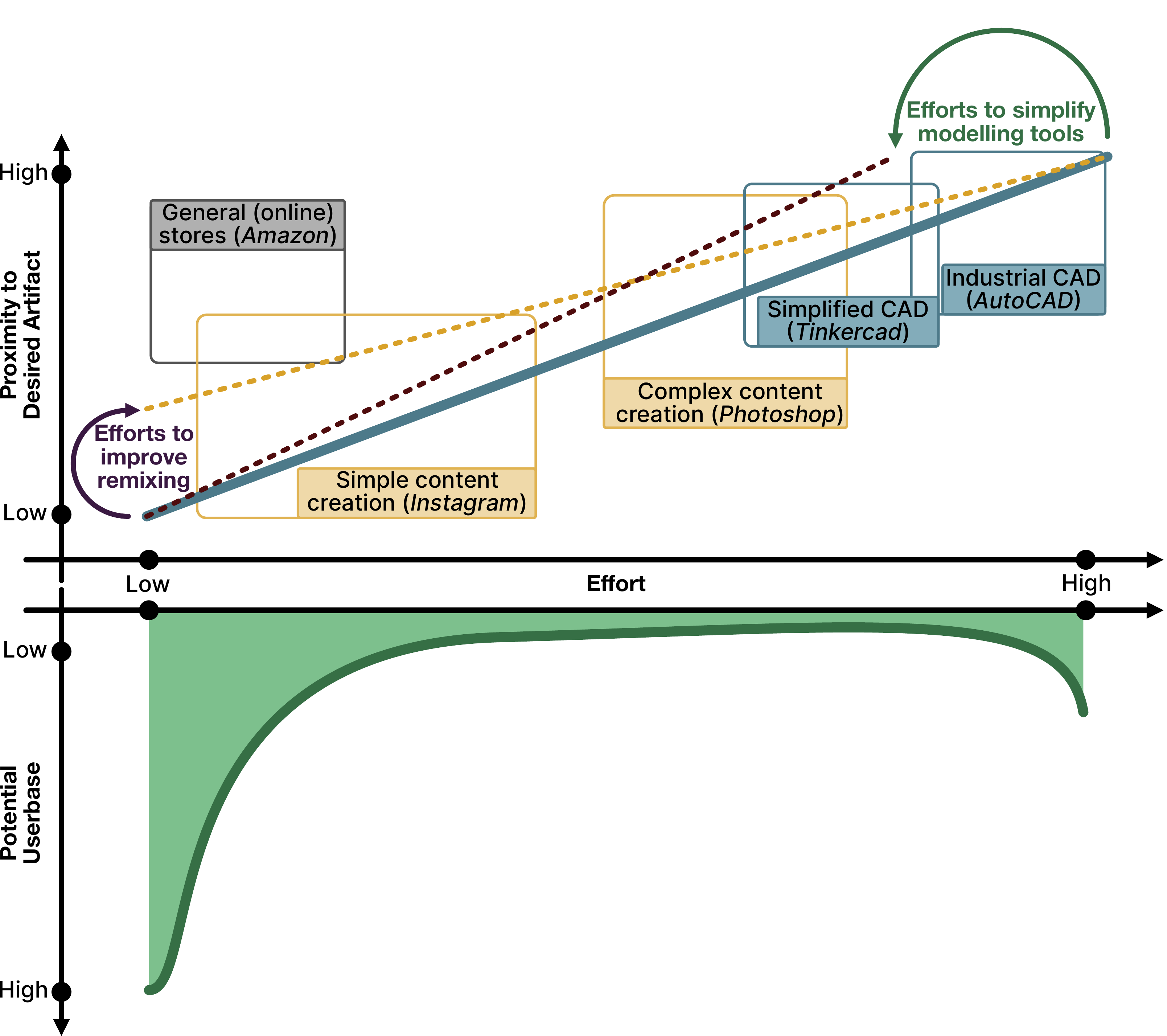}
        \caption{Simplified relation between effort (of a workflow) and the potential gap between desired and achievable artifact. Various prototypes simplify complex modeling tools (green arrow). We argue that the opposite direction (purple), based on lowest-effort interfaces for shopping and content creation, may involve the population in PF. }
        \label{fig:model1-teaser}
    \end{figure*}
    
    However, mere ownership of or access to such devices (e.g., 3D-printers) along with the software needed (e.g., CAD software), does not make a person a user.
    While increasingly more machine knowledge can be embedded in the hard- or software itself~\cite{baudischPersonalFabrication2017}, users have to \emph{precisely} express requirements for future artifacts (e.g., dimensions).
    We consider the established notion of (3D-) modeling -- defining shapes based on simple primitives such as lines or voxels --  to be a hindrance for widespread adoption of PF.
    Defining artifacts from the ground up is appropriate for domain experts or users enjoying it and possessing intrinsic motivation for the process itself, not necessarily the result~\cite{hudsonUnderstandingNewcomers3D2016}.
    While paradoxical at first, we argue that PF must provide ways for future users to benefit from intricately tailored, personal artifacts, without resorting to defining them in great detail.
    Likewise, DF has to provide simple, low-effort tools for content creation that enable users to explore the possibilities of the technology while generating quick, yet viable, results.
    Our main argument is that these low-effort tools should not be a simplified version of an expert tool which follows a creation paradigm ''from scratch'', but should rather be radically simple interfaces which omit most modeling and required expert knowledge, reducing artifact creation to as few interactions as possible.
    We propose a model to differentiate modeling and remixing. 
    We see \textit{''remixing''} as a gradient between two extremes: \textit{''getting''} (purchases in stores) and \textit{''modeling''} (designing and defining artifacts from the ground up).
    We further survey a set of recent literature in PF and categorize them within our gradient that focuses on \emph{effort} as a core dimension.

    We partially ground our argument in parallel developments that can be observed in the fields of music, video, or image editing (e.g.,  GarageBand $\Leftrightarrow$ Logic Pro, Instagram $\Leftrightarrow$ Adobe Photoshop, and TikTok $\Leftrightarrow$ Adobe Premiere).
    These facets of \emph{content creation} have non-experts in photography, videography, or music,  creating content for communities like TikTok or Soundcloud (Figure \ref{fig:model1-teaser}).
    By relying on automation and derivative work, users (unlikely to use expert tools) are enabled to explore and generate content without explicit training, high entry barriers, and with low effort.
    One of the core arguments we propose in this work, is that one reason Instagram and TikTok were able to bring content creation to the masses, is not only their associated communities, but also their radical break with the ''creation paradigm''. 
    Instead of giving users full control over the content, as done by expert tools, creation follows \textit{''getting''} and \textit{''remixing''} paradigms, where the users select from pre-defined tools and filters, which often leverage automation (e.g., face tracking for videos).
    Standardizing these processes is an apparent reduction in expressivity. 
    However, the combination of pre-defined operations still offers a sufficient variety to be able to please the needs of the individual (who often even depends on templates to realize and explore what is possible).
    Flath et al. describe the growth of Thingiverse users as coinciding with the introduction of the customizer~\cite{flathCopyTransformCombine2017} -- a way to enable novices to create remixes without the skills to model the entire artifact.
    This customization aspect enhanced a store-like interface (i.e., Thingiverse, \textit{''getting''}) with options to tailor artifacts to one's liking and is an inherently different step than the simplification of established industry-grade CAD tools (e.g., AutoCAD) to simpler ones (e.g., Tinkercad), as depicted in Figure \ref{fig:model1-teaser}.
    Based on such parallels and our survey, we synthesize concepts crucial for the further dissemination of PF as a relevant aspect of everyday scenarios: enabling derivative works, leveraging automation, crowds, and communities, along with a focus on \textit{''remixing''} and \textit{''getting''}, instead of \textit{''modeling''} by defining artifacts (or content) from the ground up.
    
    We see this work as a call to action that PF may need a -- possibly counterintuitive -- paradigm shift to reach a wider audience beyond enthusiasts and domain experts, and to become a genuinely pervasive technology.

\section{Ubiquitous Personal Fabrication}\label{sec:background}
    \begin{figure*}[tp!]
        \centering
        \includegraphics[width=\textwidth]{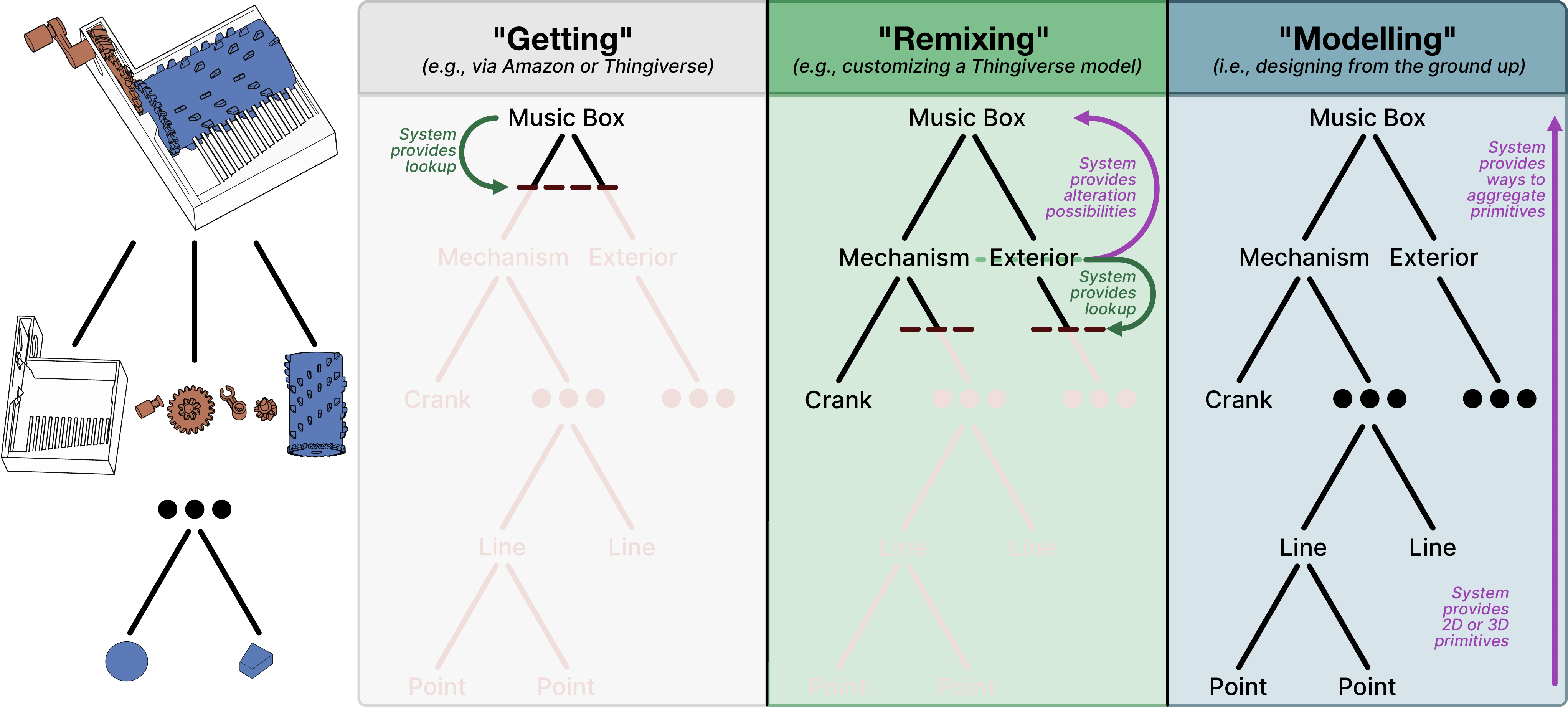}
        \caption{Compositions of a music box. The aggregation of components (nodes, leaves) to a finished artifact (root) can be used to determine the degree an artifact is remixed. Remixing (using an existing component or part) allows users to prune subtrees, thereby omitting the effort needed to define them actively (model). For the music box, systems that allow lookups for its sub-parts enable remixing through the use of existing parts. Modeling systems provide the users with primitives (2D, 3D) to compose to an artifact. If a lookup for a finished artifact succeeds, a user ''gets'' it directly. (\url{https://www.thingiverse.com/thing:53235} ''Parametric Music Box'' by wizard23 CC-BY)}
        \label{fig:remix-trees}
    \end{figure*}
    With this work, we focus on what Hudson et al. identify as ''casual makers'': people that care about results, and less about processes~\cite{hudsonUnderstandingNewcomers3D2016}, making them more akin to consumers than makers~\cite{baudischPersonalFabrication2017}.
    With them being the majority of the population, their adoption of PF will enable \textbf{ubiquitous personal fabrication} -- used for entertainment / aesthetic purposes \emph{and} functional, productive purposes, as is the case with computing.

    Non-experts in video editing became proficient generators of novel video content -- not through ever simpler manifestations of industry-grade systems like Adobe Premiere, but rather through simple tools relying on derivative works coupled with communities (e.g., TikTok).
    The same can be presumed for images (i.e., Instagram and Photoshop) or audio content (e.g., through GarageBand).
    While the importance of community and network effects can not be disregarded, the tools these platforms deliver and embed in their workflow are compelling: they rely heavily on derivative work (e.g., addition of imagery like stickers), automate previously tedious processes (e.g., through filters) and deliberately omit the majority of functions their professional counterparts provide.
    The majority of users will not generate sophisticated movie productions or photographs. %
    However, they are deeply entwined in content creation for the respective domains. %
    For PF, this may also be the case: perfectly precise, industry-grade parts may not be the aspect that drives the widespread adoption of DF.
    Mundane, easy-to-create, low-effort artifacts that still generate value (e.g., entertainment, tailored artifacts) for creators and involved communities will likely make up for a bulk of artifacts made.\\

    \textbf{\emph{Modeling}} can be defined as \textit{''to design or imitate forms''}. %
    We consider traditional 3D-modeling workflows (CAD software) to be in line with this definition: a user combining precise, fundamental geometric primitives (e.g., lines) until an object takes shape.
    When one takes existing artifacts (e.g., a model from Thingiverse) to change them, the primitive a user is working with is a finished, usable artifact.
    Alterations of it are then a \textbf{\emph{remix}}~\cite{flathCopyTransformCombine2017,alcockBarriersUsingCustomizing2016}.
    While the extremes of the gradient between \textit{''getting''} and \textit{''modeling''} are distinct (i.e., getting a complete artifact with essentially ''one click'', compared to modeling it from primitives), the transition between these two concepts is gradual, with the building blocks (i.e., primitives the users work with) progressing from fundamental shapes over their aggregations to essentially finished artifacts -- we treat this space as \textit{''remixing''}, where a \emph{degree} of modeling work is omitted through the system.
    
    To further clarify our definitions, we want to introduce a tree-analogy to emphasize the fluid transition between \textit{''getting''} over \textit{''remixing''} to \textit{''modeling''} artifacts (Figure \ref{fig:remix-trees}).
    We consider trees as they are defined in computer science for the process of defining / attaining an artifact: with a root (desired artifact), nodes (subcomponents of the artifact) and leaves (fundamental geometric primitives).
    We do not provide a formal way of creating such trees but argue that each artifact can be decomposed into them.
    The more of the nodes users have to define themselves, the closer they operate to ''modeling'' (high effort, no pruning). 
    The more nodes (and the closer they are to the root) are provided to the users, the closer they operate to \textit{''getting''} (intermediate to low effort). 
    In its simplest form, remixing is mainly \textit{''getting''} the object, the central paradigm for store interfaces (low-effort, tree pruned right below root).
    An example can be seen in Figure \ref{fig:remix-trees} for a music box. %
    It is comprised of mechanical (functional) components (a crank, gears, the cylinder defining the melody) and an enclosure. %
    
    If users are interested in \textit{\textbf{''getting''}} a (personalized) music box, they may refer to stores or repositories like Thingiverse. 
    They provide a \emph{lookup} for the particular object the users are searching for, and are able to deliver it to them.
    Thereby, users are enabled to prune all subtrees below the finished artifact and do not have to invest further work or occupy themselves with subtrees.
    This is possible, as long as the lookup succeeded or the user accepts an alternative artifact.
    We deliberately consider online shopping to be a feasible ''branch'' of PF where design and manufacturing have happened already without user intervention and may surpass a user's personal acceptance threshold (i.e., the product fits ''good enough'').
    To achieve a higher degree of personalization, users may consider \textit{\textbf{''remixing''}} a personalized music box.
    For instance, by downloading a parametric design from Thingiverse and customizing it. %
    A set of subtrees can be pruned, while the remaining subtrees may require modeling work from the users.
    Alternatively, they may require input of parameters for a generative design (e.g., for embossed text). %
    To achieve the highest degree of control and personalization, users may resort to \textit{\textbf{''modeling''}} the music box.
    This requires the use of more sophisticated software that provides primitives to aggregate (e.g., via CSG, constructive solid geometry) to more complex parts.
    Users then combine 2D features like lines and rectangles to 3D features, which in turn are combined to higher-level components, like gears.
    Some systems provide both 2D and 3D primitives to reduce workload (allowing users to prune leaves).
    The more work is omitted in the tree (i.e., through a lookup), the more one can consider a process to be \textit{''getting''}.
    If an interface is able to deliver an artifact directly, one may consider it an interface for \textit{''getting''} objects.
    In contrast, if a user is required to aggregate a majority of components by combining primitives, one may consider the interface to follow a \textit{''modeling''} approach.\\

\section{Design Tools for Personal Fabrication}
    \begin{figure*}[ht!]
        \hspace*{-5em}\includegraphics[width=1.145\linewidth]{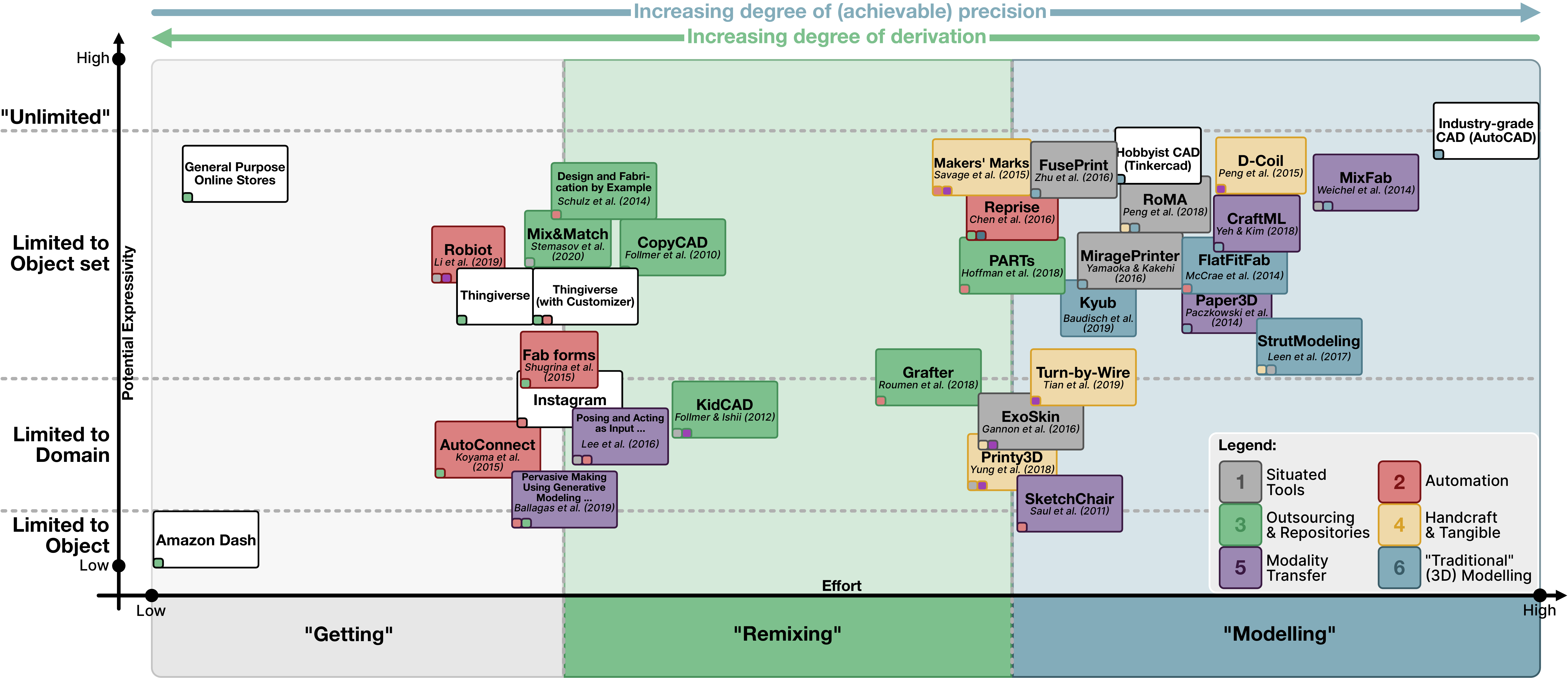}
        \caption{A selection of tools arranged by effort and potential expressivity. We decompose effort into three ranges from 'getting' over 'remixing' to 'modeling'. Expressivity is similarly segmented, ranging from tools meant for a single artifact, tools for specific domains to tools limited to entire object ranges.
        We classify systems by their \textbf{core approach} to artifact definition (background) and secondary traits (corner).}
        \label{fig:model1-main}
    \end{figure*}
    
    To emphasize these two lines of approaches currently explored in research (\textit{''modeling''} -- \textit{''remixing''}) and the established paradigm of \textit{''getting''} artifacts, we selected papers related to simplifying the definition of artifacts from the last 10 years from venues such as CHI, UIST or TOG. %
    The initial set consisted of $73$ papers, which were distilled to $27$ works used to illustrate the gradient.
    Our selection is not a holistic overview of the field, but consists of papers that highlight unique aspects and approaches of the community to PF. %
    The core focus is on the \emph{design} of artifacts, and less their \emph{fabrication}.
    While fabrication itself poses intricate challenges, it is likely the least ''personal'' aspect of PF -- ideally to be automated and optimized without user intervention to reduce effort.
    Systems allowing overlap between fabrication and design~\cite{pengRoMAInteractiveFabrication2018,yamaokaMiragePrinterInteractiveFabrication2016,tianTurnbyWireComputationallyMediated2019} or augmenting handcraft~\cite{pengDCoilHandsonApproach2015} remain part of the classification, as they can be considered an abstraction of CAD tools and their complexities.\\

    As mentioned in our motivation, \emph{effort} needed to achieve a satisfactory result is the core dimension we consider relevant for ubiquitous adoption of PF.
    Ideally, the more effort users invest in a modeling process, the closer they should get to their envisioned artifact.
    However, this relation is not linear or increasing monotonously: low-effort interfaces exist both for artifact acquisition (e.g., online shopping) and content creation (e.g., Instagram).
    Therefore, Figure \ref{fig:model1-main} abstracts the ''proximity to desired artifact (c.f., Figure \ref{fig:model1-teaser}) to a general notion of ''expressivity'' -- a subjective measure of the possibilities a tool enables. %
    
    A tool with low expressivity is able to deliver one single artifact -- an example is the Amazon Dash button, which is pre-configured to \textit{''get''} a specific product, chosen once before. 
    A step above such systems in terms of expressivity, are systems limited to a specific domain. %
    For example, systems for customizing glasses~\cite{ballagasExploringPervasiveMaking2019} or  chairs~\cite{saulSketchChairAllinoneChair2011}, reduce the effort needed, by constraining the result domain of the process.
    Tools exclusively intended to remix toys~\cite{follmerKidCADDigitallyRemixing2012} are domain-specific, while tools intended to remix any static shape are more expressive.
    Tools that additionally account for dynamics/kinematics can be considered as even more expressive. 
    
    We consider the potential expressivity of industry-grade CAD tools to be close to potentially unlimited, as they allow the design of highly complex artifacts (e.g., entire engines), along with their assembly and simulation.
    While not omnipotent yet, we consider them to be on the verge of having no set limits to their expressivity -- \emph{as long as appropriate \textbf{effort} is invested}.\\
    
    The method with the least effort needed has the highest potential degree of adoption, when considering the \emph{entire} population.
    Low-effort means low friction, low entry barriers, with users quickly being able to succeed with their tasks.
    Examples for such an interface are the aforementioned Amazon Dash buttons, as most steps to acquire an (non-personalized) artifact are omitted.
    A step above such a lowest-effort interface are general interfaces.
    General, ''all-purpose'' stores cover a high expressivity, while specialized ones cover a lower degree thereof.
    Thingiverse, can be regarded as a specialized store, bound by the current limits of DF hardware.
    If augmented with customization options, store interfaces inherit aspects of remix- or tailoring-oriented interfaces.
    This mainly refers to parametric designs, as found for instance on Thingiverse, but also services that provide customers with the ability to tailor product lines as they desire (e.g., furniture). 
    The more the customization options, the more effort the users may have to invest. %
    We consider interfaces that embrace derivative works to exhibit low to intermediate effort, relative to established modeling paradigms.
    In PF, these are interfaces to repositories like Thingiverse~\cite{stemasovMixMatchOmitting2020}, or interfaces for customizing parametric designs~\cite{shugrinaFabFormsCustomizable2015}.
    The definition of artifacts ''from scratch'' can be considered to be a high-effort activity (with potentially high reward: a one-of-a-kind artifact).
    Primitives are aggregated and combined with operations (e.g., CSG) and, with increasing detail, yield an exact rendition of the desired artifact.
    (3D) modeling can be considered a dominant paradigm used in products and research prototypes to define arbitrary artifacts. %

    Based on the retrieved set of literature (Figure \ref{fig:model1-main}), we derived 6 approaches to reduce effort for PF:
    tools that are 
    1) situated,
    2) automation-supported, 
    3) repository-based, 
    4) handcraft- or tangibility-oriented, 
    5) employing modality transfer %
    and lastly 6) largely ''traditional'' modeling tools.
    The horizontal arrangement of the works in Figure \ref{fig:model1-main} is based on the degree of effort required and the degree of derivation a tool employs. 
    The vertical arrangement is based on the expressivity of the tools or their applicability to different artifact classes.
    The positioning of the works inside a sector was derived by mutual comparisons with respect to their required effort and achieved expressivity.
    The following paragraphs introduce systems that are representative for these approaches.
    
    \textit{\textbf{Situated tools (1)}} are an approach to bridge the disconnect between the space a future artifact is meant to reside in, and the space it is being specified in.
    This allows easier embedding of real features \cite{zhuFusePrintDIY5D2016,stemasovMixMatchOmitting2020,yamaokaMiragePrinterInteractiveFabrication2016}, especially when they might be hard to measure and digitize~\cite{gannonExoSkinOnBodyFabrication2016}.
    These aspects reduce effort in comparison to more disconnected methods for modeling or remixing, while allowing users to preview their work before or during fabrication~\cite{pengRoMAInteractiveFabrication2018,stemasovMixMatchOmitting2020}.
    Situated tools can be either modeling tools incorporating the real world as reference (e.g., \cite{weichelMixFabMixedrealityEnvironment2014,yamaokaMiragePrinterInteractiveFabrication2016}, pruning subtrees of real-world features), or be remixing tools embracing entire outsourced artifacts~\cite{follmerCopyCADRemixingPhysical2010,stemasovMixMatchOmitting2020} (pruning at or close to the root).
    Situated tools provide the advantage of the correct physical \emph{context}, thereby omitting steps like measurements, but may suffer from limited expressivity due to alternative input and output devices, compared to established 3D modeling workflows.
    
    \textit{\textbf{Automation (2)}} enables users to omit specific steps after providing input to a design tool.
    Systems relying on generative design can infer models from sketches~\cite{saulSketchChairAllinoneChair2011} or movement~\cite{liRobiotDesignTool2019}.
    They are also able to transfer modalities like speech (i.e., descriptions of a product) to a personalized design~\cite{ballagasExploringPervasiveMaking2019}.
    Parametrized systems expose a limited set of dimensions for users to explore and customize~\cite{shugrinaFabFormsCustomizable2015}.
    They may also rely on existing physical objects as input to generate additional geometry~\cite{koyamaAutoConnectComputationalDesign2015,chenRepriseDesignTool2016}.
    This class of systems either enriches modeling tools (pruning subtrees) or enables users to interact with simple interfaces to object remixing~\cite{shugrinaFabFormsCustomizable2015} or generation~\cite{liRobiotDesignTool2019} (pruning close to the root).
    By automating steps, previously complex procedures to define geometry are omitted.
    However, current automation approaches are constrained by the parametrized or previously learned geometries and can be limited to certain predefined sets of objects.
    
    Works focusing on \textit{\textbf{Repositories and Outsourcing (3)}} aim to benefit from finished models or features~\cite{follmerCopyCADRemixingPhysical2010,hofmannGreaterSumIts2018,stemasovMixMatchOmitting2020} at the expense of expressivity of the workflow, by including them in a design process.
    Systems may either rely on features existing in the immediate surroundings of the user (e.g., \cite{follmerCopyCADRemixingPhysical2010,follmerKidCADDigitallyRemixing2012,stemasovMixMatchOmitting2020}) or objects found in crowd-based model repositories (e.g., \cite{hofmannGreaterSumIts2018,roumenGrafterRemixing3DPrinted2018,stemasovMixMatchOmitting2020}).
    Outsourcing may also happen through automation (work offloaded to the system) or by relying on curated, centralized repositories (e.g., \cite{schulzDesignFabricationExample2014}) or pre-defined parametric templates. %
    Most such systems can be assigned to the range of \textit{''remixing''} but may either require low effort~\cite{follmerCopyCADRemixingPhysical2010} or are meant for more complex domains requiring more modeling-like procedures~\cite{roumenGrafterRemixing3DPrinted2018}.
    Tools that embrace outsourced work are generally limited to the databases of artifacts they rely on -- both in terms of object diversity and how much is encoded in the objects (e.g., static geometry or parametrized geometry). They furthermore need robust retrieval methods to benefit the user.
        
    Systems relying on \textit{\textbf{Handcraft and tangible tools (4)}} enable users to interact with their design in a more immediate and tangible fashion.
    By providing tangible, pre-defined components, users may not only manipulate them directly, but also allow a system to replace placeholders with complex geometry~\cite{savageMakersMarksPhysical2015}, allowing users to omit their design.
    This enables a more unconstrained interaction with the design materials~\cite{yungPrinty3DInsituTangible2018}.
    Interactive fabrication is enabled by mediated input to the fabrication device~\cite{tianTurnbyWireComputationallyMediated2019} or augmenting previously manual processes with computerized support~\cite{pengDCoilHandsonApproach2015}. 
    This likewise enables users to rely on existing features in their designs, or ease primitive-based modeling processes. %
    Most such systems rely on modeling as a paradigm -- however, they are closer to its original definition, grounded in handcraft.
    With generative aspects~\cite{savageMakersMarksPhysical2015} and reliance on simpler building blocks~\cite{yungPrinty3DInsituTangible2018}, some venture close to remixing-like procedures.
    While tangible tools re-introduce feedback lost with most digital fabrication techniques, they likewise re-introduce a skill and learning curve, thereby increasing effort in some cases.

    \textit{\textbf{Modality Transfer (5)}} is likewise a common occurrence in literature.
    By avoiding traditional CAD metaphors (e.g., aggregation of primitives), systems enable novices to express their requirements.
    Sketching-based systems enable users to omit precision in their design process, which is either not necessary to achieve a functional artifact~\cite{saulSketchChairAllinoneChair2011}, or is inferred by a system relying on artificial intelligence~\cite{liRobiotDesignTool2019}.
    Examples include the use of metaphors known from natural materials~\cite{paczkowskiPaper3DBringingCasual2014}, gestures~\cite{weichelMixFabMixedrealityEnvironment2014,leePosingActingInput2016}, speech~\cite{ballagasExploringPervasiveMaking2019} or programming~\cite{yehCraftML3DModeling2018}.
    Some systems that employ a modality transfer, still employ modeling as their core paradigm~\cite{yehCraftML3DModeling2018,weichelMixFabMixedrealityEnvironment2014,paczkowskiPaper3DBringingCasual2014}, while others radically omit explicit steps to define them from the ground up~\cite{ballagasExploringPervasiveMaking2019,leePosingActingInput2016}.
    Tools employing modality transfer \emph{without} omitting modeling processes generally reduce effort with respect to learning, but less so with respect to effort during the process itself. 
    Approaches that combine novel modalities with low-effort processes, reduce effort throughout the entire design process~\cite{ballagasExploringPervasiveMaking2019,leePosingActingInput2016}.
    
    \textit{\textbf{''Traditional'' Modeling tools (6)}} are approaches that, at their core, aim to simplify modeling as a paradigm, thereby making it more accessible to novices.
    This includes the reduction of available primitives~\cite{baudischKyub3DEditor2019}, operations~\cite{mccraeFlatFitFabInteractiveModeling2014}, or fidelity~\cite{leenStrutModelingLowFidelityConstruction2017}.
    While they generally achieve a simpler (and often less expressive) process than industry-grade toolchains, they rely on the paradigm of modeling, which, while the most expressive, requires effort for the complete definition of artifacts nonetheless.\\
    
    All previously presented approaches have in common that they reduce the required effort for PF.
    However, a majority of them still employs modeling-like approaches. %
    Prototypes employing high degrees of automation~\cite{ballagasExploringPervasiveMaking2019,koyamaAutoConnectComputationalDesign2015}) or ones that outsource modeling work to crowds~\cite{roumenGrafterRemixing3DPrinted2018,stemasovMixMatchOmitting2020}, venture close to \emph{''modeling-free personal fabrication''} and \textit{''getting''}.
    They enable users to quickly and often effortlessly receive uniquely personalized results without having to define them from the ground up. 
    These low-effort / high-expressivity procedures are crucial for personal digital fabrication, both for productive, functional artifacts, but also for content creation as such.
    DF can be a synthesis of mere artifact acquisition (e.g., online shopping) and content creation (e.g., for platforms like Instagram).
    If appropriate, low-effort procedures are provided to users, DF itself may become a ubiquitous and inclusive endeavor for users currently not involved in it.

\section{Conclusion -- A Call to Action}\label{sec:conclusion}
    With this work, we wanted to emphasize that personal fabrication research may and should focus on ways to circumvent work and effort needed to achieve one's goals.
    We argue that if PF is meant to be adopted by a wide range of users, the \textbf{benefits} (e.g., a personalized product) do not dictate the \textbf{process} (e.g., 3D modeling).
    The bulk of objects designed with the means of \emph{personal} fabrication will likely not exhibit a complexity that would dictate precise and lengthy processes.
    We argue that both researchers and practitioners should consider finding novel ways to \textbf{omit} processes akin to modeling, instead of aiming to simplify them further and further. 
    This is likely a crucial component to achieve widespread adoption in households, instead of fablabs and other (technology-)enthusiasts' spaces, making PF as ubiquitous and relevant as personal computing itself became over time.
    A second, crucial component to widespread adoption of PF is the embedding thereof into less serious content creation contexts, linked to vibrant communities.
    Our main argument is that these low-effort tools should not be a simplified version of an expert tool, but rather radically simple interfaces which omit most of the process and required expert knowledge, reducing artifact creation to as few interactions as possible -- \textit{''getting''}.

    Based on our survey, we derived specific aspects that may support system-driven HCI research to achieve, or venture towards modeling-free personal fabrication: a focus on derivative works leveraging automation, crowds, and communities, with a general direction towards \textit{''remixing''} and \textit{''getting''}, instead of \textit{''modeling''}.

    Pervasive PF can not ignore the environmental implications.
    One important future challenge for pervasive PF will be the potential environmental impact. 
    Physical artifacts can not be as easily removed as mass-produced digital content such as images. 
    Therefore, our call for action is also directed to sustainability research that addresses this issue. 
    We argue that the path towards mass usage is a realistic one (with many researchers finding ways to enable and accelerate PF for novices) and, therefore, the question of sustainability has to be addressed now, \emph{before it becomes ubiquitous}.
    
    For an actual novice in DF, the decision is not between Tinkercad and AutoCAD, but rather between ''Product A'' and ''Product B'' online, due to the low-effort experience of \textit{''getting''}.
    The inclusion of a majority of people in personal DF requires different approaches than are currently dominant in research.
    Low-effort approaches to DF will likely be the ones to enable ubiquitous personal fabrication.

    \bibliographystyle{tufte}
    \bibliography{zotero}
    
    \vfill
    \begin{IEEEbiographynophoto}{Evgeny Stemasov}
    is a second-year PhD student at Ulm University in Germany. 
    He is interested in personal fabrication along with the design and engineering processes enabling it for novices and experienced users alike. 
    Evgeny holds a Master's Degree in Media Computer Science from Ulm University. Contact him at evgeny.stemasov@uni-ulm.de
    \end{IEEEbiographynophoto}
    
        \begin{IEEEbiographynophoto}{Enrico Rukzio}
    is full professor in the Institute of Media Informatics at Ulm University, Germany.
    He is interested in designing intelligent interactive systems that enable people to be more efficient, satisfied and expressive in their daily lives.
    Enrico received his PhD in computer science from the University of Munich. Contact him at enrico.rukzio@uni-ulm.de
    \end{IEEEbiographynophoto}
    
    \begin{IEEEbiographynophoto}{Jan Gugenheimer}
    is Assistant Professor for Computer Science at Télécom Paris (Institut Polytechnique de Paris) in the DIVA group.
    He is working on several topics around Mixed Reality (Augmented Reality and Virtual Reality) with focus on Human-Computer Interaction. 
    Jan received his PhD in computer science from Ulm University. Contact him at jan.gugenheimer@telecom-paris.fr
    \end{IEEEbiographynophoto}

\end{document}